\documentclass[aps,superscriptaddress,prc,twocolumn,nofootinbib]{revtex4}

\usepackage{graphicx}
\usepackage{epstopdf}
\usepackage{subfigure}
\usepackage{epsfig}
\usepackage{amsmath,amssymb,amsfonts}
\usepackage{color}
\usepackage[utf8]{inputenc}
\allowdisplaybreaks
\usepackage[bookmarks,
                bookmarksopen = true,
                bookmarksnumbered = true,
                linktocpage,
                colorlinks = true,
                linkcolor = blue,
                urlcolor  = blue,
                citecolor = blue,
                anchorcolor = green,
                hyperindex = true,
                hyperfigures]
                {hyperref}
\usepackage{multirow}
\usepackage{array}

\begin{document}

\title{Flavor hierarchy of parton energy loss in quark-gluon plasma from a Bayesian analysis}

\author{Wen-Jing Xing}
\affiliation{Institute of Frontier and Interdisciplinary Science, Shandong University, Qingdao, Shandong 266237, China}
\affiliation{Institute of Particle Physics and Key Laboratory of Quark and Lepton Physics (MOE), Central China Normal University, Wuhan, 430079, China}

\author{Shanshan Cao}
\email{shanshan.cao@sdu.edu.cn}
\affiliation{Institute of Frontier and Interdisciplinary Science, Shandong University, Qingdao, Shandong 266237, China}

\author{Guang-You Qin}
\email{guangyou.qin@mail.ccnu.edu.cn}
\affiliation{Institute of Particle Physics and Key Laboratory of Quark and Lepton Physics (MOE), Central China Normal University, Wuhan, 430079, China}

\date{\today}

%%%%%%%%%%%%%%%%%%%%%%%%%%%%%%%%%%%%%%%%%%%%%%%%%%%%%%%%%%%%%%%%%%%%%

\begin{abstract}

The quenching of light and heavy flavor hadrons in relativistic heavy-ion collisions probes the color and flavor dependences of parton energy loss through a color-deconfined quark-gluon plasma (QGP), and thus offers an important test of QCD-based calculation at extremely high density and temperature. By combining a next-to-leading order perturbative QCD calculation of parton production, a general ansatz of parton energy loss functions and parton fragmentation functions, we calculate the nuclear modification of various hadron species -- charged hadrons, $D$ mesons and $B$-decayed $J/\psi$ -- over a wide transverse momentum regime. Comparing our calculations to the  experimental data using the Bayesian statistical analysis, we perform a first simultaneous extraction of the energy loss functions of gluons ($g$), light quarks ($q$), charm quarks ($c$) and bottom quarks ($b$) inside the QGP. We find that the average parton energy loss at high energies follows the expected hierarchy of $\langle \Delta E_g \rangle > \langle \Delta E_q \rangle \sim \langle \Delta E_c \rangle > \langle \Delta E_b \rangle$, while the parton energy loss distribution can further test the QCD calculations of parton interaction with the dense nuclear matter.
We also find that the reduction of  experimental uncertainties can significantly improve the precision of the extracted parton energy loss functions inside the QGP.

\end{abstract}

\maketitle

%%%%%%%%%%%%%%%%%%%%%%%%%%%%%%%%%%%%%%%%%%%%%%%%%%%%%%%%%%%%%%%%%%%%%

\section{Introduction}
\label{sec:Introduction}

Jets serve as a valuable probe of the nuclear matter created in relativistic heavy-ion collisions. The suppression of the high transverse momentum ($p_{\rm T}$) jet and hadron spectra in nucleus-nucleus (A+A) collisions compared to those in proton-proton (p+p) collisions, known as jet quenching, is recognized as smoking-gun evidence for the formation of a color deconfined QCD medium -- quark-gluon plasma (QGP) -- in these energetic A+A collisions~\cite{Wang:1992qdg}. Jet quenching originates from both elastic scatterings~\cite{Bjorken:1982tu,Braaten:1991we,Djordjevic:2006tw,Qin:2007rn} and inelastic scatterings~\cite{Baier:1994bd,Baier:1996kr,Zakharov:1996fv,Gyulassy:1999zd,Wiedemann:2000za,Arnold:2002ja,Wang:2001ifa} experienced by jet partons as they propagate through the QGP. With a wealth of experimental data on jet observables, studies of jet-medium interactions have been extended from the quenching of inclusive hadron and jet spectra~\cite{ATLAS:2014ipv,CMS:2016xef,ALICE:2018vuu,Bass:2008rv,He:2018xjv,Xing:2019xae,Qiu:2019sfj,JETSCAPE:2022jer} to nuclear modification of dihadron and dijet asymmetry~\cite{ATLAS:2010isq,Zhang:2007ja,Qin:2010mn,Chen:2016vem}, correlations between photon (or $Z$ boson) and its triggered hadron (or jet)~\cite{CMS:2012ytf,Qin:2009bk,Chen:2017zte,Luo:2018pto,Zhang:2018urd}, intra-structures of jets~\cite{CMS:2012nro,CMS:2013lhm,ATLAS:2014dtd,STAR:2021kjt,Chang:2016gjp,Tachibana:2017syd,Chang:2019sae,JETSCAPE:2023hqn} and hadron chemistry within jets~\cite{Luo:2021voy,Chen:2021rrp,Sirimanna:2022zje}. A full understanding of these observables involves not only the medium modification of jets~\cite{Majumder:2010qh,Qin:2015srf,Blaizot:2015lma}, but also jet-induced medium excitation~\cite{Cao:2020wlm,Cao:2022odi}.

One of the central goals of studying jets in heavy-ion collisions is utilizing them to extract the properties of nuclear matter at different temperature scales. For instance, considerable efforts have been devoted to the extraction of the jet transport coefficient $\hat{q}$ in both hot and cold nuclear matter~\cite{JET:2013cls,Andres:2016iys,Ru:2019qvz,JETSCAPE:2021ehl,Xie:2022ght}. This $\hat{q}$ parameter measures the transverse momentum broadening of an energetic parton due to its elastic scatterings with the medium~\cite{Baier:2002tc,Majumder:2008zg}. It is directly related to the density of medium constituents, and is a key quantity in estimating the intensity of the medium-induced gluon bremsstrahlung from the hard parton~\cite{Guo:2000nz,Majumder:2009ge,Sirimanna:2021sqx}.
Recently, it has been proposed that the enhancement of $\hat{q}$ near the critical endpoint may also be applied to probing the QCD phase diagram~\cite{Wu:2022vbu}. Besides $\hat{q}$, one may also directly extract the amount of parton energy loss ($\Delta E$) inside the QGP. For example, the energy loss distribution (or the quenching weight)  was obtained in Ref.~\citep{Arleo:2017ntr} by comparing a simplified model calculation with the experimental data on the nuclear modification factor ($R_\mathrm{AA}$) of single inclusive hadrons. In this work, the form of the parton energy loss is taken from the BDMPS medium-induced gluon spectrum~\cite{Baier:1996kr,Baier:1996sk,Arleo:2002kh}, and both light flavor hadrons and prompt $J/\psi$'s are assumed to be produced only from gluon fragmentation while $D$ mesons are only from charm quark fragmentation. Later in Ref.~\cite{He:2018gks}, a more general ansatz of the jet energy loss distribution is assumed and convoluted with the medium-modified jet function, based on which the flavor-averaged jet energy loss function is extracted using the jet $R_\mathrm{AA}$ data with the Bayesian statistic interference. This methodology is then extended in Ref.~\citep{Zhang:2022rby} for constraining the energy loss distribution functions of both gluons and charm quarks from the nuclear modification of $J/\psi$ at high $p_\mathrm{T}$. However, a simultaneous data-driven constraint on the energy loss functions of all parton species -- gluon ($g$), light quarks ($q=u,d,s$), charm quark ($c$) and bottom quark ($b$) -- is still absent in literature. This is the focus of our present study, which is essential in unraveling the color, flavor and mass hierarchy of parton energy loss.

Considering the successful perturbative QCD (pQCD) description of different hadron species at high $p_\mathrm{T}$ ($\gtrsim10$~GeV) in p+p collisions, we will convolute the well-established next-to-leading-order (NLO) perturbative calculations of parton production~\cite{Aversa:1988vb,Jager:2002xm}, a general ansatz of parton energy loss inside the QGP, and parton fragmentation~\cite{Kretzer:2000yf,Kneesch:2007ey,Kniehl:2007erq}. By comparing the nuclear modification of charged hadrons, $D$ mesons and $B$-decayed $J/\psi$ between our model calculation and the experimental data at the Large Hadron Collider (LHC)~\citep{CMS:2016xef,CMS:2017qjw,CMS:2017uuv}, we will perform a simultaneous extraction of the energy loss functions of $g$, $q$, $c$ and $b$ using the state-of-the-art Bayesian analysis method. The mean values of the parton energy loss at high energies exhibit a clear flavor hierarchy, $\langle \Delta E_g \rangle > \langle \Delta E_q \rangle \sim \langle \Delta E_c \rangle > \langle \Delta E_b \rangle$, and their distribution functions can provide a stricter test on QCD calculations on parton scatterings inside a hot nuclear matter. The uncertainties propagated from the experimental data to our result will also be explored.

\section{Hadron production and medium modification}
\label{sec:JQ_framework}

The differential cross section for high-$p_\mathrm{T}$ hadron production in p+p collisions can be factorized as
\begin{align}
\label{eq:dsigma_pp}
\frac{d \sigma_{\mathrm{pp}\to hX}}{ dp_\mathrm{T}^h} &= \sum_j \int dp_\mathrm{T}^j  dz \frac{d \hat{\sigma}_{\mathrm{pp}\to jX}}{dp_\mathrm{T}^j}(p_\mathrm{T}^j)
\nonumber\\
& \times D_{j\to h}(z)  \delta\left(p_\mathrm{T}^h -z p_\mathrm{T}^j\right).
\end{align}
In the above equation, $d\hat{\sigma}_{\mathrm{pp}\to jX}$ represents the cross section for inclusive parton ($j$) production and can be calculated by convoluting the parton distribution functions (PDFs) of the two colliding protons with the partonic scattering cross section; $D_{j\to h}$ denotes the fragmentation function (FF) of parton $j$ to a given hadron species $h$. In this work, the PDFs are taken from the CTEQ parameterizations~\cite{Pumplin:2002vw}, the partonic cross section is evaluated perturbatively at the NLO~\cite{Aversa:1988vb,Jager:2002xm}, and the FFs are taken from Ref.~\cite{Kretzer:2000yf} for charged hadrons, Ref.~\cite{Kneesch:2007ey} for $D$ mesons, and Ref.~\cite{Kniehl:2007erq} for $B$ mesons. As shown in our earlier studies~\citep{Xing:2019xae,Xing:2021bsc,Liu:2021izt}, this combination provides a good description of charged hadron, $D$ meson and $B$ meson spectra at $p_\mathrm{T}\gtrsim 10$~GeV in p+p collisions at the LHC energy. Note that within the NLO framework, one can consistently include both quark and gluon fragmentations to heavy and light flavor hadron productions. In particular, the gluon fragmentation dominates the charged hadron production at $p_{\rm T}<50$~GeV, and contributes to almost 40\% $D$ meson and 50\% $B$ meson productions up to $p_{\rm T}$ beyond 50~GeV.

In heavy-ion collisions, we assume the parton production from the initial hard scatterings is a superposition of $\langle N_\mathrm{coll} \rangle$ p+p collisions, where $\langle N_\mathrm{coll} \rangle$ represents the average number of (binary) nucleon-nucleon collisions in each A+A collision. The nuclear shadowing effect is taken into account by modifying the nucleon PDFs using the EPS09 parameterizations~\cite{Eskola:2009uj}. The hard partons then interact with the QGP and lose transverse momentum $\Delta p_\mathrm{T}$ according to an energy loss function $W_\mathrm{AA}(x)$~\cite{He:2018gks}, where $x=\Delta p_\mathrm{T} / \left\langle \Delta p_\mathrm{T}\right\rangle$ is the ratio between the transverse momentum loss in a particular event and its mean value. Hadronization is assumed to take place outside the medium, and therefore the vacuum fragmentation function can be applied as in p+p collisions. With this setup, the binary-collision-number-rescaled cross section of hadron production reads:
%\begin{align}
%\label{eq:dsigma_AA}
%\frac{1}{\langle N_\mathrm{coll} \rangle} & \frac{d \sigma_{\mathrm{AA}\to hX}}{ dp_\mathrm{T}^h} = \sum_j \int dp_\mathrm{T}^j  dx dz \frac{d \hat{\sigma}_{\mathrm{p'p'}\to jX}}{dp_\mathrm{T}^j}(p_\mathrm{T}^j) \nonumber\\
%\times & W_\mathrm{AA}(x) D_{j\to h}(z)
%\delta\left(p_\mathrm{T}^h -z(p_\mathrm{T}^j-x\langle \Delta p_\mathrm{T}^j\rangle)\right).
%\end{align}
\begin{align}
\label{eq:dsigma_AA}
\frac{1}{\langle N_\mathrm{coll} \rangle} &\frac{d \sigma_{\mathrm{AA}\to hX}}{ dp_\mathrm{T}^h} = \sum_j \int_{0}^{\infty} dp_\mathrm{T}^j  \int_{0}^\frac{p_\mathrm{T}^j}{\left\langle \Delta p_\mathrm{T}^j \right\rangle } dx \int_{0}^{1} dz
\nonumber\\
\times & \frac{d \hat{\sigma}_{\mathrm{p'p'}\to jX}}{dp_\mathrm{T}^j}(p_\mathrm{T}^j) W_\mathrm{AA}(x) D_{j\to h}(z)
\nonumber\\
\times & \delta\left(p_\mathrm{T}^h -z(p_\mathrm{T}^j-x\langle \Delta p_\mathrm{T}^j\rangle)\right).
\end{align}
Here, $d\hat{\sigma}_{\mathrm{p'p'}\to jX}$ represents the parton cross section after including the shadowing effect, and different species of hard partons lose different amount of $\langle\Delta p_\mathrm{T}\rangle$ which is parameterized as
\begin{equation}
\label{eq:averageLoss}
\left\langle \Delta p_\mathrm{T}^j \right\rangle = C_j \beta_g p_\mathrm{T}^{\gamma} \mathrm{log}(p_\mathrm{T}),
\end{equation}
where $\beta_g$ controls the overall magnitude of gluon energy loss, $\gamma$ tunes its $p_\mathrm{T}$ dependence, and $C_j$ represents the parton energy loss ratio relative to the gluon's -- 1 for gluon and $C_q, C_c, C_b$ for light, charm and bottom quarks respectively. The normalized distribution function of parton energy loss is given by
\begin{eqnarray}
\label{eq:WAA}
W_\mathrm{AA}(x)=\frac{\alpha^{\alpha} x^{\alpha-1}e^{-\alpha x} }{\Gamma(\alpha)},
\end{eqnarray}
with $\alpha$ being a model parameter, which in principle can be computed from jet energy loss theory. Generally, parameters $\gamma$ and $\alpha$ above can also depend on the parton flavor. However, they were shown to be consistent between gluon and charm quark in an earlier study using the $J/\psi$ data~\citep{Zhang:2022rby}. Thus in the present work, they are set to be the same for $g$, $q$, $c$ and $b$. Releasing this assumption will consume much longer computational time, and will be left for our future effort. Thus, we have in total 6 parameters -- $\beta_g, C_q, C_c, C_b, \gamma$ and $\alpha$ -- in this analysis. In this work, we do not include a specific energy loss model, while uncertainties in our extracted parameters can arise from the nuclear PDFs and FFs we use. The energy loss function we obtain in the end also depends on the parameterization scheme designed here. However, different schemes should lead to comparable results if they capture the main feature of the parton energy loss and meanwhile provide good descriptions of the experimental data after the Bayesian calibrations.

The ratio of Eq.~(\ref{eq:dsigma_AA}) to Eq.~(\ref{eq:dsigma_pp}) gives the nuclear modification factor ($R_\mathrm{AA}$) of a given type of hadron.
By comparing our model calculation to the $R_\mathrm{AA}$ data of charged hadrons, $D$ mesons and $B$-decayed $J/\psi$, we will extract the above 6 parameters and obtain the energy loss functions of different parton species. Note that in order to exclude the effects of non-perturbative interactions of partons with the QGP at low $p_\mathrm{T}$~\cite{He:2014cla,Liu:2017qah,Xing:2021xwc}, and the coalescence process during their hadronization~\cite{Fries:2003vb,Cao:2019iqs} at low to intermediate $p_\mathrm{T}$, we restrict our current study to the high $p_\mathrm{T}$ ($>10$~GeV) regime where the perturbative calculation is reliable.
Instead of directly using the $B$ meson $R_\mathrm{AA}$ to constrain the energy loss function of $b$ quarks, we decay $B$ mesons to $J/\psi$ through Pythia~\citep{Sjostrand:2006za}, given that the experimental uncertainties of the latter is much smaller than the former~\cite{CMS:2017uoy,CMS:2017uuv}.

\section{Calibration with Bayesian analysis}
\label{sec:Bayesian_analysis}

The Bayesian inference method has been successfully employed in both soft and hard sectors of heavy-ion physics, such as the extractions of initial condition, bulk transport coefficients and the equation of state of the QGP medium~\citep{Novak:2013bqa,Pratt:2015zsa,Bernhard:2015hxa,Bernhard:2016tnd,Auvinen:2017fjw,Bernhard:2019bmu,JETSCAPE:2020shq}, jet transport coefficient $\hat{q}$~\citep{JETSCAPE:2021ehl}, heavy quark diffusion coefficient $D_\mathrm{s}$~\citep{Xu:2017obm}, and the energy loss distribution of hard partons and jets~\citep{He:2018gks,Zhang:2022rby}. It has also been utilized to constrain the fluctuating structure of protons with experimental data on coherent and inherent diffractive $J/\psi$ production in electron-proton collisions at HERA~\citep{Mantysaari:2022ffw}. In this work, we implement the Bayesian inference method  to simultaneously constrain the energy loss functions of different partons ($g$, $q$, $c$ and $b$) through the QGP medium.

Our model calibration on the parameter set, denoted by $\boldsymbol\theta$, is based on the following Bayes' theorem,
\begin{eqnarray}
\label{eq:Bayes_theorem}
P\left( \boldsymbol\theta | \mathrm{data} \right) \propto P\left( \boldsymbol\theta \right) P\left( \mathrm{data}|\boldsymbol\theta \right),
\end{eqnarray}
where $P\left( \boldsymbol\theta | \mathrm{data} \right)$ on the left represents the posterior distribution of the parameter set  given the knowledge of the experimental data, $P\left(\boldsymbol\theta \right)$ on the right represents the prior distribution of $\boldsymbol\theta$ without any knowledge of data, and $P\left( \mathrm{data}|\boldsymbol\theta \right)$ measures the likelihood of a given set of $\boldsymbol\theta$ by comparing its model output with the data. In the present study, $\boldsymbol\theta=(\beta_g, C_q, C_c, C_b, \gamma,\alpha)$ is a 6-dimensional vector, and the likelihood function is assumed to follow the Gaussian form as
\begin{equation}
\label{eq:Gaussian_likelihood}
P\left( \mathrm{data}|\boldsymbol\theta \right) = \prod_i \frac{1}{\sqrt{2 \pi} \sigma_i} e^{- \left[y_i(\boldsymbol\theta) - y_i^\mathrm{exp}\right]^2 \big/ (2 \sigma_i^2)},
\end{equation}
where $y_i(\boldsymbol\theta)$ is the model output at a given data point $i$, and $y_i^\mathrm{exp}$ and $\sigma_i$ are respectively the mean value and the error of the experimental observation at the point $i$. The statistic error and the systematic error of the experimental data are combined in our analysis.

Using Eqs.~(\ref{eq:Bayes_theorem}) and~(\ref{eq:Gaussian_likelihood}), we perform a Markov-Chain Monte-Carlo (MCMC) random walk~\cite{MCMC} in the parameter space according to the Metropolis-Hasting algorithm. Each step of this random walk updates the parameter set based on its current location in the parameter space. We start from a random location and execute $10^7$ steps for the Markov-Chain to reach equilibrium, after which the locations given by further steps can constituent equilibrium distributions of the model parameters, which are the posterior distributions we seek in this work. We generate another $10^7$ steps after reaching equilibrium, from which we draw these posterior distributions. To exclude correlation between adjacent steps, we draw one sample of $\boldsymbol\theta$ from every 10 steps, and use $10^6$ samples in the end to produce the probability distribution of the parameter space.

A uniform prior distribution $P\left( \boldsymbol\theta \right)$ is assumed in the regions of $\beta_g \in (0,10)$, $C_q \in (0, 1)$, $C_c \in (0,1)$, $C_b \in (0,1)$, $\gamma \in (-0.15, 0.5)$ and $\alpha \in (0,15)$, which are sufficiently wide for our model output to cover all the experimental data. In order to accelerate the calibration speed, a Gaussian Process (GP) emulator~\cite{GPE1,GPE2} is applied for the model-to-data comparison when we scan through the 6-dimensional parameter space. This emulator is trained with our model results on 600 design points uniformly distributed in the prior range of $\boldsymbol\theta$ above, and then used as a fast surrogate of the real perturbative calculation during the Bayesian analysis on the hadron $R_\mathrm{AA}$.

\begin{figure}[tbp]
\centering
\includegraphics[width=1.0\linewidth]{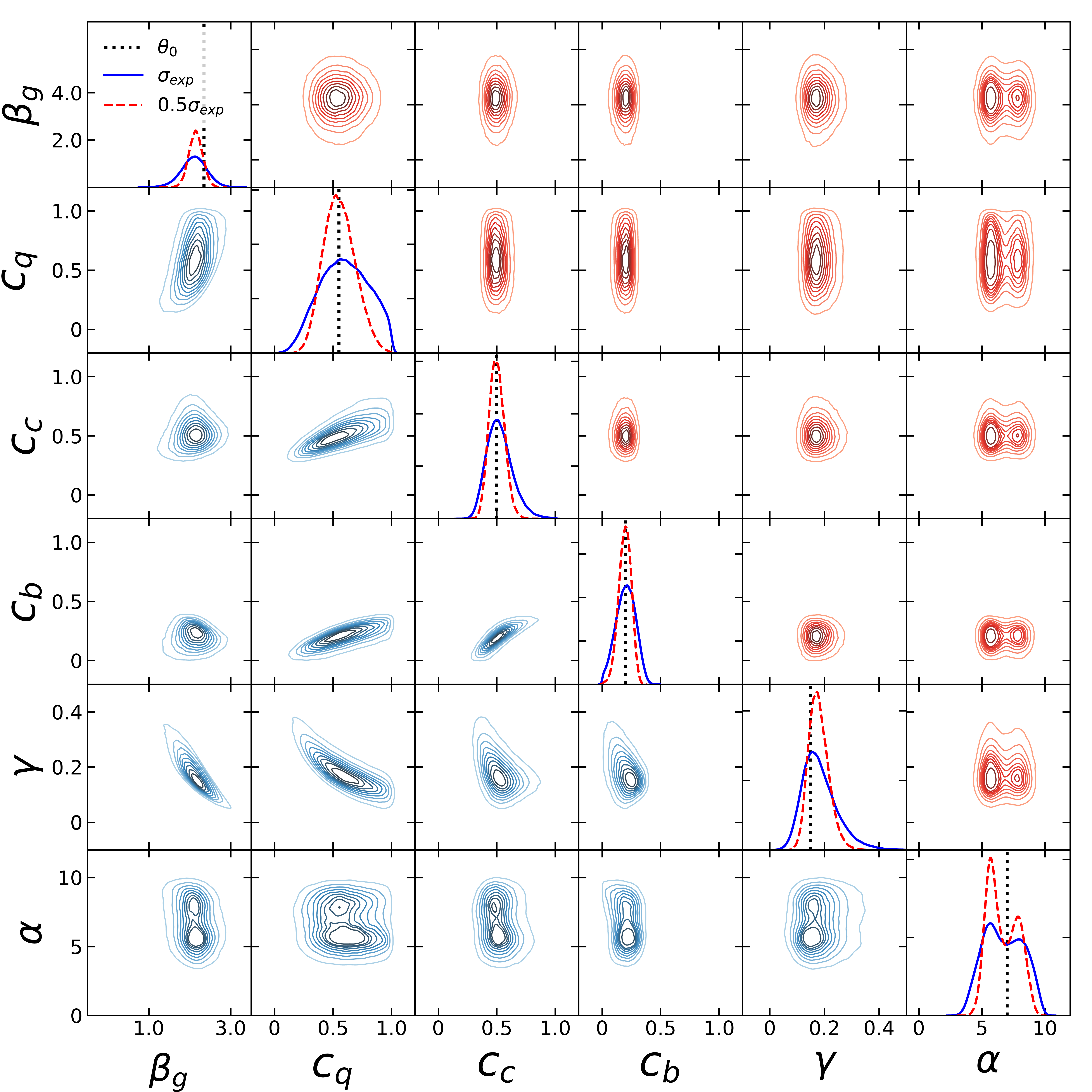}
\caption{(Color online) Closure test: posterior distributions of model parameters (diagonal panels) and their correlations (off-diagonal panels) extracted from pseudo-experimental data on the $R_\mathrm{AA}$'s of charged hadrons, $D$ mesons and $B$-decayed $J/\psi$, with black vertical lines denoting the input (``ground truth") values of $\boldsymbol\theta_0 = (2.35, 0.55, 0.5, 0.2, 0.15, 7)$ used to generate the pseudo-data. The lower triangle (blue) is obtained using the original error bars of the experimental data~\cite{CMS:2016xef,CMS:2017qjw,CMS:2017uuv} while the upper triangle (red) is obtained using halved error bars.}
\label{fig:closure_test}
\end{figure}

To validate our setup of the GP emulator and the Bayesian analysis framework, we first conduct a closure test in Fig.~\ref{fig:closure_test} based on a set of pseudo-experimental data generated by the model calculation. We start with a particular design point -- $\boldsymbol\theta_0 = (2.35, 0.55, 0.5, 0.2, 0.15, 7)$ -- as indicated by the black vertical lines in the figure, and use it to calculate the $R_\mathrm{AA}$ of charged hadrons, $D$ mesons and $B$-decayed $J/\psi$ with the perturbative method developed in Sec.~\ref{sec:JQ_framework}. These model results are used to replace the mean values of the experimental data in 0-10\% Pb+Pb collision at $\sqrt{s_\mathrm{NN}}=5.02$~TeV~\citep{CMS:2016xef,CMS:2017qjw,CMS:2017uuv}, while the error bars of these data points are taken from the original CMS measurements. Using these pseudo-data, we then examine whether the posterior distribution of $\boldsymbol\theta$ extracted from the Bayesian analysis agrees with the ``ground-truth" value $\boldsymbol\theta_0$ that we implant. The posterior distributions of the 6 model parameters are presented in Fig.~\ref{fig:closure_test}, with diagonal panels showing the probability distributions of individual parameters and off-diagonal ones showing the correlations between different pairs of parameters. The lower triangle (blue) is obtained by using the original error bars of the CMS data in the Bayesian analysis, while the upper triangle (red) is obtained by using halved error bars. Meanwhile, we present the 90\% credible regions (C.R.'s) of the extracted model parameters in Tab.~\ref{tab:closure_test}, which are obtained by discarding the lowest 5\% and highest 5\% areas of their distribution functions. One can see that the probability distribution, or the 90\% C.R., of each model parameter can be constrained to its pre-set ``true" value within this analysis framework. Halving the error bars of our pseudo-data narrows the 90\% C.R.'s of the posterior distributions, and thus improves the precision of the extracted parameters.

\begin{table}[ht!]
\begin{center}
  \begin{tabular}{p{1.5cm}<{\centering}|p{1.5cm}<{\centering}|p{2.5cm}<{\centering}|p{2.5cm}<{\centering}}
   \hline
                         &   $\boldsymbol\theta_0$          &   with $\sigma_\mathrm{exp}$          & with $0.5\sigma_\mathrm{exp}$    \\ \hline
    $\beta_g$    &    2.35   &    (1.565, 2.614)   &  (1.862, 2.49)    \\ \hline
    $C_q$          &    0.55   &    (0.266, 0.928)  & (0.344, 0.789)   \\ \hline
    $C_c$          &    0.5     &    (0.362, 0.725)   &  (0.398, 0.61)    \\ \hline
    $C_b$          &    0.2     &    (0.063, 0.331)   & (0.102, 0.278)   \\ \hline
    $\gamma$   &    0.15   &    (0.095, 0.303)   &  (0.125, 0.245)    \\ \hline
    $\alpha$      &    7.0     &    (4.349, 9.146)   & (5.01, 8.561)   \\ \hline
  \end{tabular}
\end{center}
  \caption{The 90\% C.R.'s of the model parameters extracted from the pseudo-experimental data, the third column from using the original uncertainties of the experimental data, and the fourth column from using the halved uncertainties.}
   \label{tab:closure_test}
\end{table}

\section{Extracting parton energy loss functions}
\label{sec:Results}

Using the validated statistical analysis framework above, we now extract the distributions of parton energy loss from the real $R_\mathrm{AA}$ data of charged hadrons, $D$ mesons and $B$-decayed $J/\psi$ in 0-10\% Pb+Pb collisions at $\sqrt{s_\mathrm{NN}}=5.02$~TeV~\citep{CMS:2016xef,CMS:2017qjw,CMS:2017uuv}.

\begin{figure}[tp]
\centering
\includegraphics[width=1.0\linewidth]{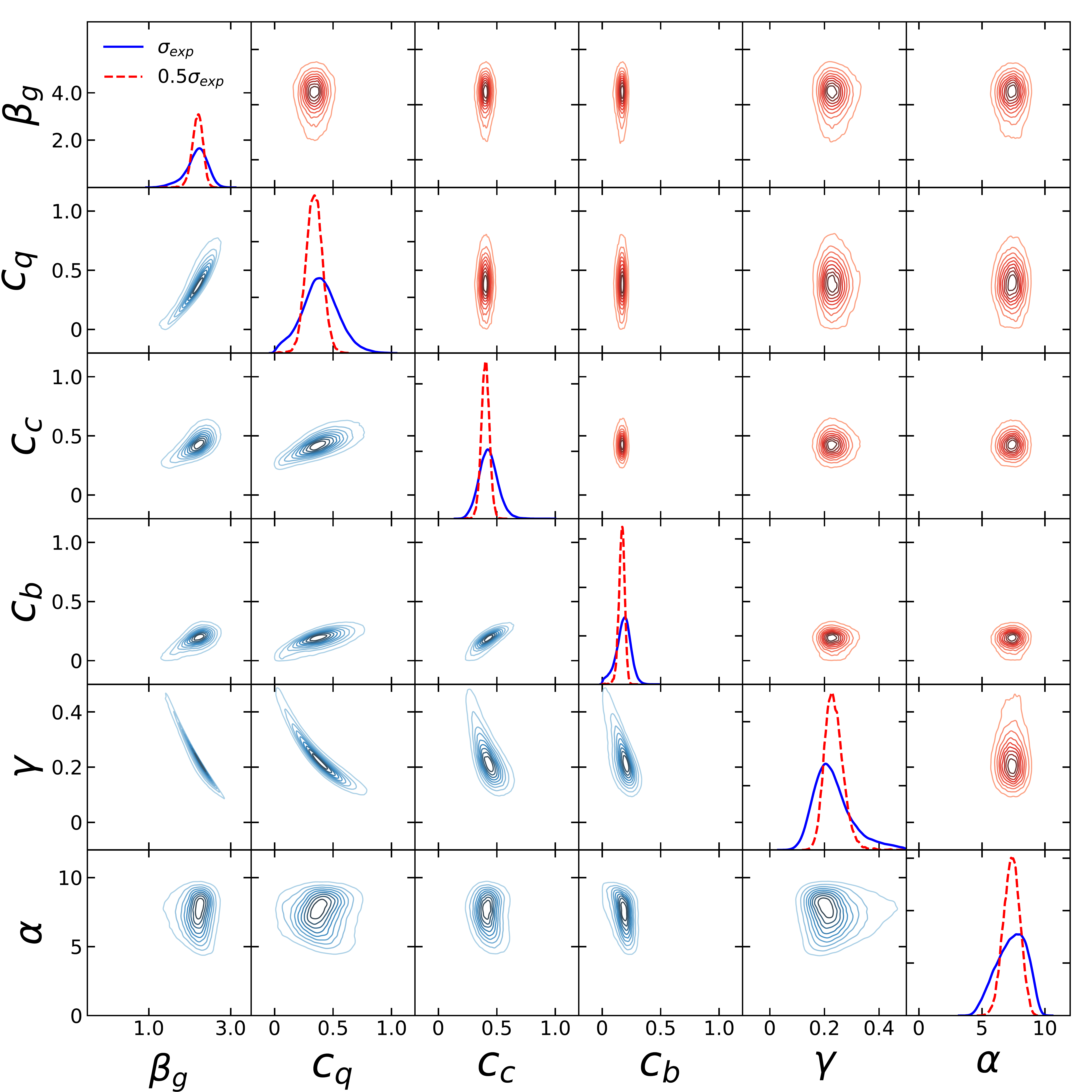}
\caption{(Color online) Posterior distributions of model parameters (diagonal panels) and their correlations (off-diagonal panels) extracted from the CMS data on the $R_\mathrm{AA}$'s of charged hadrons, $D$ mesons and $B$-decayed $J/\psi$ in 0-10\% Pb+Pb collisions at $\sqrt{s_\mathrm{NN}}=5.02$~TeV~\citep{CMS:2016xef,CMS:2017qjw,CMS:2017uuv}. The lower triangle (blue) is obtained using the original error bars of the data while the upper triangle (red) is obtained using halved error bars.}
\label{fig:posterior_correlation}
\end{figure}

\begin{table}[ht!]
\begin{center}
  \begin{tabular}{p{2.5cm}<{\centering}|p{2.5cm}<{\centering}|p{2.5cm}<{\centering}}
   \hline
                         &  with  $\sigma_\mathrm{exp}$         &  with 0.5$\sigma_\mathrm{exp}$    \\ \hline
    $\beta_g$    &    (1.646, 2.56)   &  (1.96, 2.39)    \\ \hline
    $C_q$          &    (0.129, 0.65)   & (0.226, 0.454)   \\ \hline
    $C_c$          &    (0.3, 0.567)   &  (0.344, 0.459)    \\ \hline
    $C_b$          &  (0.065, 0.277)   & (0.124, 0.207)   \\ \hline
    $\gamma$   &    (0.137, 0.378)  &  (0.184, 0.295)    \\ \hline
    $\alpha$      &    (5.287, 9.061)  & (6.266, 8.401)   \\ \hline
  \end{tabular}
\end{center}
  \caption{The 90\% C.R.'s of the model parameters extracted from the CMS data on the $R_\mathrm{AA}$'s of charged hadrons, $D$ mesons and $B$-decayed $J/\psi$ in 0-10\% Pb+Pb collisions at $\sqrt{s_\mathrm{NN}}=5.02$~TeV~\citep{CMS:2016xef,CMS:2017qjw,CMS:2017uuv}, middle column from using the original uncertainties of the experimental data, and right column from using the halved uncertainties. }
  \label{table:parameters}
\end{table}

In Fig.~\ref{fig:posterior_correlation}, we present the posterior probability distributions of the 6 model parameters extracted from the Bayesian fit to the experimental data. The diagonal panels show the probability distributions of individual parameters while the off-diagonal ones show the two-parameter joint distributions that quantify their correlations. The lower triangle (blue) is from using the original error bars of the experimental data in our Bayesian analysis, while the upper triangle (red) from the halved error bars. The 90\% C.R.'s of these parameters are summarized in Tab.~\ref{table:parameters}, including results from using both the full (middle column) and halved (right column) error bars. One can see that the model parameters can be well constrained by comparing our perturbative calculation to the experimental data. An inverse correlation between $\beta_g$ (or $C_q$, $C_c$, $C_b$) and $\gamma$ is observed. This is consistent with earlier findings in Refs.~\cite{He:2018gks,Zhang:2022rby}, and can be understood with Eq.~(\ref{eq:averageLoss}) where they both contribute positively to the parton energy loss. The posterior distribution of the parameter $\alpha$ here is also consistent with that obtained earlier in Ref.~\citep{Zhang:2022rby}. In addition, reducing the uncertainties from the experimental data makes the posterior distributions of the model parameters narrower, thus helps better constrain the parton energy loss functions. Note that the ranges of $C_q$ obtained here are consistent with the ratio of the Casimir factors between quark and gluon.

\begin{figure}[tbp!]
\centering
\includegraphics[width=0.97\linewidth]{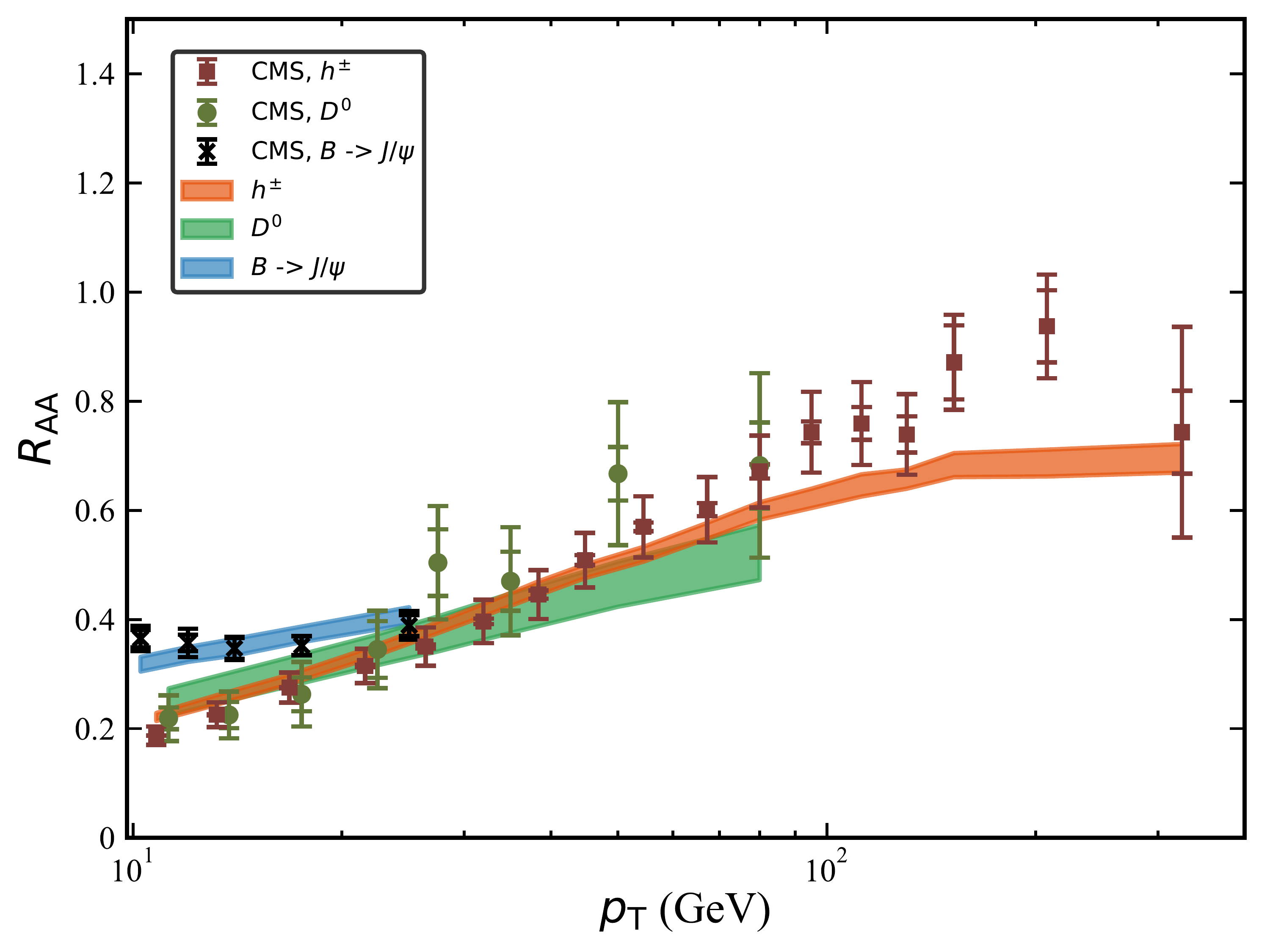}
\caption{(Color online) The nuclear modification factors of charged hadrons, $D$ mesons and $B$-decayed $J/\psi$ in 0-10\% Pb+Pb collisions at $\sqrt{s_\mathrm{NN}}=5.02$~TeV, compared between the perturbative calculation using the posterior distributions of model parameters and the CMS data~\cite{CMS:2016xef,CMS:2017qjw,CMS:2017uuv}. The theoretical bands correspond to the $\pm 1 \sigma$ uncertainties around the mean values.}
\label{fig:posterior_RAA}
\end{figure}

Shown in Fig.~\ref{fig:posterior_RAA} is our model result after the Bayesian calibration. We first calculate the $R_\mathrm{AA}$'s of different hadrons using parameters drawn from their posterior distributions. Then, we evaluate the mean values and standard deviations ($\sigma$) of $R_\mathrm{AA}$'s at each $p_\mathrm{T}$, and present their $\pm 1\sigma$ bands in Fig.~\ref{fig:posterior_RAA}. Our calibrated model calculation provides a simultaneous description on the nuclear modification factors of charged hadrons, $D$ mesons and $B$-decayed $J/\psi$ measured by the CMS Collaboration.

\begin{figure}[tbp!]
\centering
\includegraphics[width=0.97\linewidth]{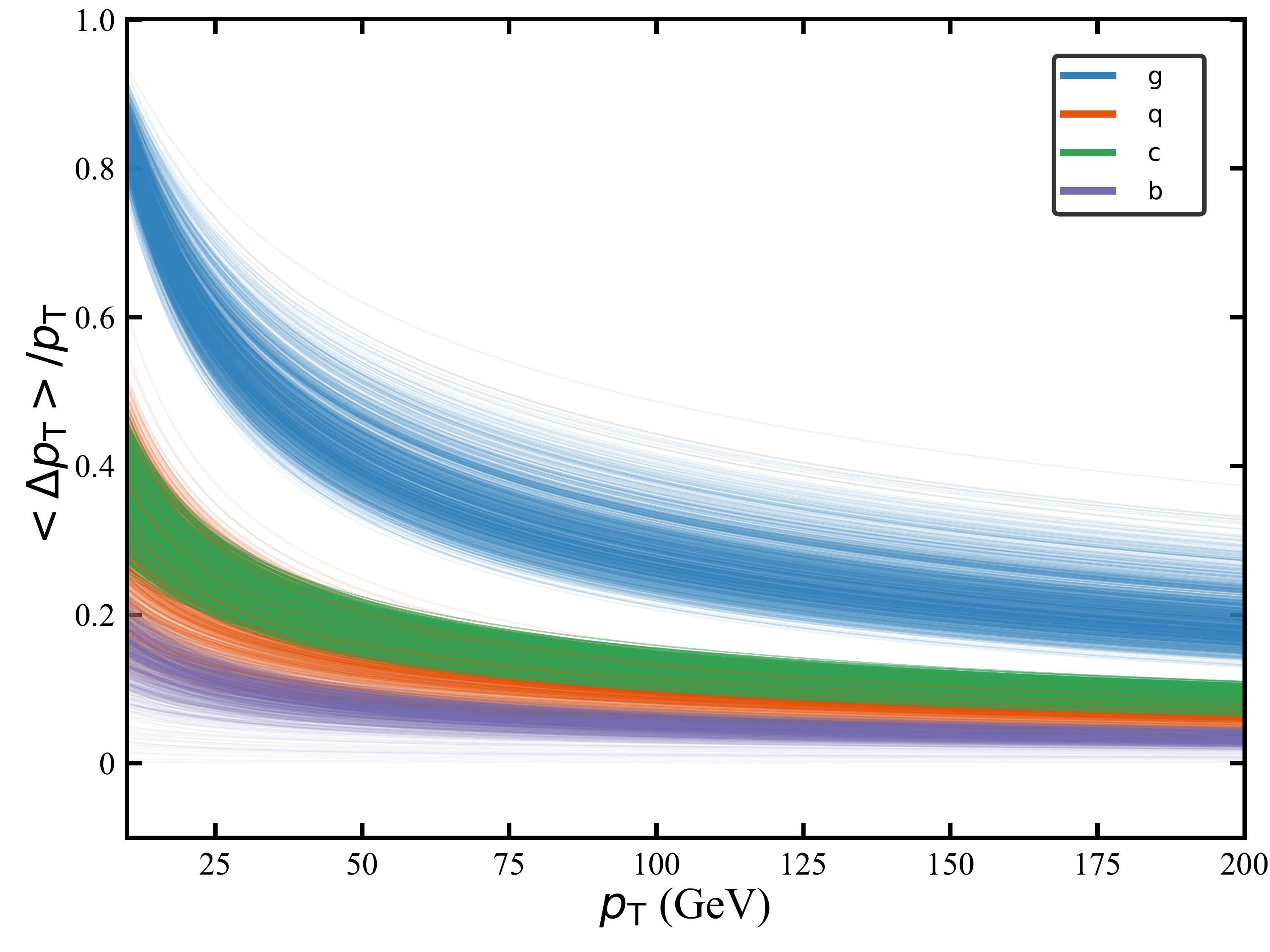}
\caption{(Color online) The average fractional transverse momentum loss of different species of partons in 0-10\% Pb+Pb collisions at $\sqrt{s_\mathrm{NN}}=5.02$~TeV.}
\label{fig:posterior_parton_eloss}
\end{figure}

\begin{figure}[tbp!]
\centering
\includegraphics[width=0.97\linewidth]{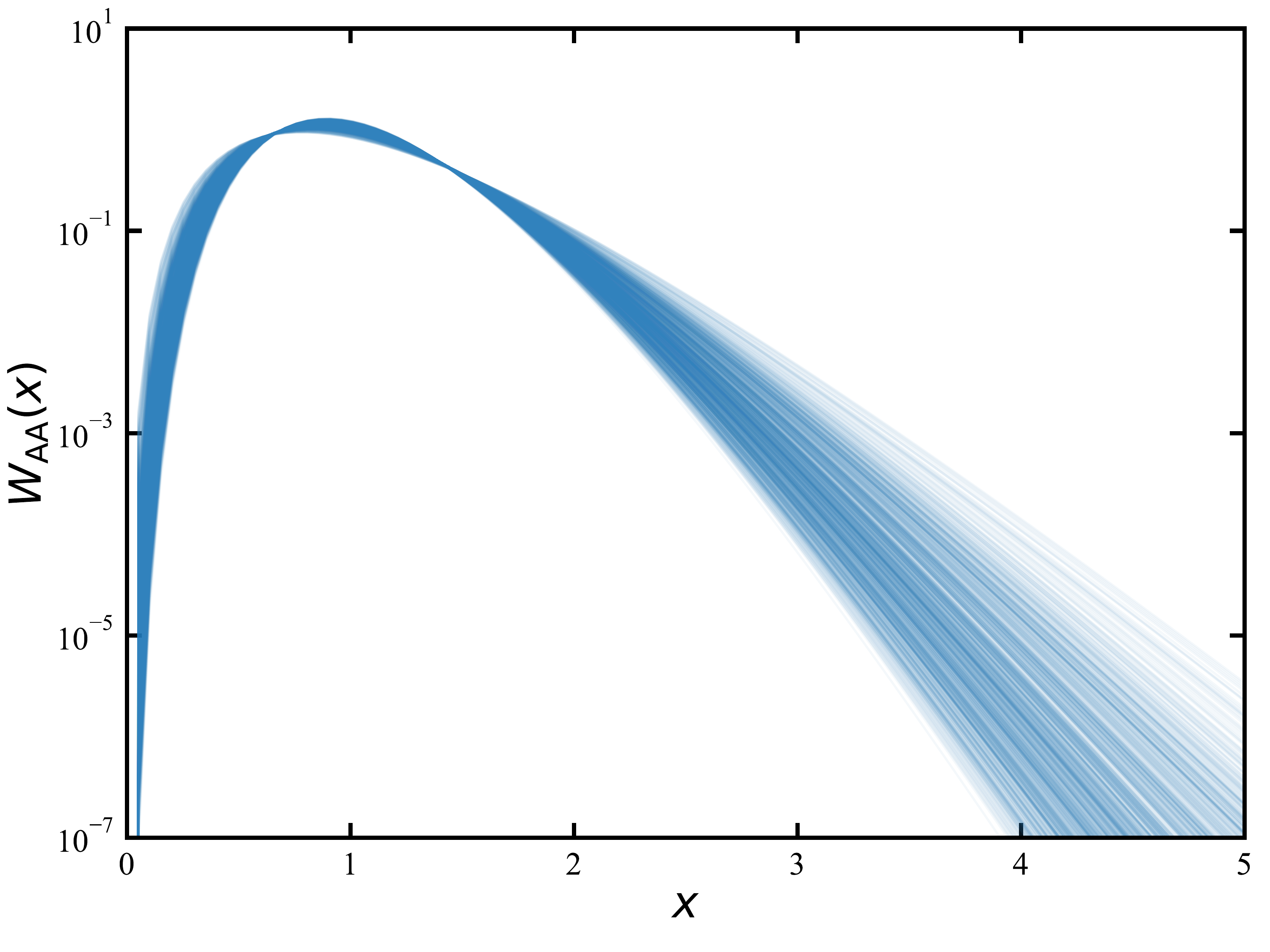}
\caption{(Color online) The distribution of parton energy loss in 0-10\% Pb+Pb collisions at $\sqrt{s_\mathrm{NN}}=5.02$~TeV.}
\label{fig:posterior_parton_eloss_dist}
\end{figure}

Using the parameters constrained by the $\pm 1\sigma$ band of the $R_\mathrm{AA}$ results above, we generate the average fractional transverse momentum loss ($\langle \Delta p_\mathrm{T} \rangle/p_\mathrm{T}$) according to Eq.~(\ref{eq:averageLoss}), and compare results between $g$, $q$, $c$ and $b$ in Fig.~\ref{fig:posterior_parton_eloss}. From the figure, one can clearly observe the flavor hierarchy of parton energy loss through the QGP: gluon loses about twice energy than quark does due to the larger color factor of the former. Light quarks and charm quark also suffer stronger energy loss than bottom quark does due to the larger mass of the bottom. Since we focus on the high $p_\mathrm{T}$ regime where the perturbative calculation is reliable, the mass effect on the charm quark energy loss is small. As a result, the extracted values of $\langle \Delta p_\mathrm{T} \rangle/p_\mathrm{T}$ are similar between light and charm quarks. Such flavor hierarchy of parton energy loss from our data-driven analysis is consistent with expectation from perturbative QCD calculations on both elastic and inelastic parton energy loss inside the QGP~\cite{Zhang:2003wk,Cao:2017hhk}. In the end, we present the distribution function of parton energy loss in Fig.~\ref{fig:posterior_parton_eloss_dist}. This distribution can in principle be calculated from jet quenching theory, thus can provide a stringent test on the QCD calculations of parton interactions with the QGP.

\section{Summary}
\label{sec:summary}

We have conducted a systematic data-driven analysis on the parton energy loss inside a QGP medium. A perturbative framework has been developed for calculating the nuclear modification of high $p_\mathrm{T}$ hadron production in relativistic heavy-ion collisions, which convolutes the cross section of parton production at NLO, a general ansatz of the parton energy loss function inside the QGP, and parton fragmentation functions into different hadrons. By comparing the nuclear modification factors of charged hadrons, $D$ mesons and $B$-decayed $J/\psi$ between our calculation and the LHC data, we have performed a first simultaneous extraction of energy loss of gluon, light quarks, charm quark and bottom quark using the Bayesian interference method. Our result shows a clear flavor hierarchy of parton energy loss at high energies, $\langle \Delta E_g \rangle > \langle \Delta E_q \rangle \sim \langle \Delta E_c \rangle > \langle \Delta E_b \rangle$ inside a hot nuclear matter, consistent with perturbative QCD expectation. We also find that a reduction of the data uncertainties can significantly improve the precision of the extracted parton energy loss functions. Our current study provides a model-independent method to quantitatively constrain the flavor dependence of parton energy loss from the experimental data. The extracted distribution function of parton energy loss also provides a stringent constraint on the perturbative QCD calculation of parton interactions with the QGP.

\section*{Acknowledgments}

We are grateful to helpful discussions with Xiang-Yu Wu, Feng-Lei Liu, Weiyao Ke and Shan-Liang Zhang. This work was supported by the National Natural Science Foundation of China (NSFC) under Grant Nos.~12175122, 2021-867, 12225503, 11890710, 11890711, and 11935007. Some of the calculations were performed in the Nuclear Science Computing Center at Central China Normal University (NSC$^3$), Wuhan, Hubei, China.

\bibliographystyle{h-physrev5}
\bibliography{SCrefs}

\end{document}